\documentclass[a4paper]{jpconf}%
\usepackage{graphicx}%
\usepackage{amsmath}%
\usepackage{bm}%
\usepackage{sidecap}%
\usepackage{pdflscape}%
\usepackage{textgreek}%

\begin{document}%

\title{Phase-Field Study of Polycrystalline Growth and Texture Selection During Melt Pool Solidification}

\author{Rouhollah Tavakoli and Damien Tourret}
\address{IMDEA Materials, Getafe, Madrid, \sc Spain}

\ead{rouhollah.tavakoli@imdea.org, damien.tourret@imdea.org}

\begin{abstract}%
Grain growth competition during solidification determines microstructural features, such as dendritic arm spacings, segregation pattern, and grain texture, which have a key impact on the final mechanical properties. During metal additive manufacturing (AM), these features are highly sensitive to manufacturing conditions, such as laser power and scanning speed. The melt pool (MP) geometry is also expected to have a strong influence on microstructure selection. Here, taking advantage of a computationally efficient multi-GPU implementation of a quantitative phase-field model, we use two-dimensional cross-section simulations of a shrinking MP during metal AM, at the scale of the full MP, in order to explore the resulting mechanisms of grain growth competition and texture selection. We explore MPs of different aspect ratios, different initial (substrate) grain densities, and repeat each simulation several times with different random grain distributions and orientations along the fusion line in order to obtain a statistically relevant picture of grain texture selection mechanisms.
Our results show a transition from a weak to a strong $\langle10\rangle$ texture when the aspect ratio of the melt pool deviates from unity.
This is attributed to the shape and directions of thermal gradients during solidification, and seems more pronounced in the case of wide melt pools than in the case of a deeper one.
The texture transition was not found to notably depend upon the initial grain density along the fusion line from which the melt pool solidifies epitaxially.%
\end{abstract}%

\section{Introduction}%

The competition of columnar grains during directional
solidification of metallic alloys occurs in many industrial
processes like welding and additive manufacturing. It may
promote the formation of a specific texture, and consequently the
anisotropy in the mechanical properties of final products.
In addition to melt convection and crystal nucleation ahead of the
solid-liquid interface, the interactions between columnar grains
mainly depend on the applied thermal field and physical properties
of the material. According to the classical theory by Walton and Chalmers \cite{walton1959origin}, dendritic
grains with their preferred dendritic growth direction better aligned with the temperature gradient are favored during the
growth competition. While this theory was applied to a wide range
of applications with a reasonable outcome, experimental
works \cite{d2002morphological,wagner2004grain,zhou2008mechanism}
revealed that growth competition mechanisms are more complicated in
practice. Recent phase-field (PF) studies on bi-crystal growth competition
under a one-dimensional linear temperature field
\cite{li2012phase,tourret2015growth,takaki2016two}, have provided
major insight in the fundamental understanding of grain growth
competition mechanisms. However, simulations of dendritic grain growth
competition in non one-dimensional temperature fields, with the level of
accuracy afforded by PF, remain scarce \cite{elahi2022multiscale,elahi2023grain}.
Here, we use parallel quantitative PF modeling to study columnar grain
growth competition under a more general thermal condition.
For this purpose, we consider the epitaxial polycrystalline
solidification in the transverse cross-section of a melt pool
corresponding to the selective laser melting (SLM) of a Nickel-based alloy. In particular, we explore the effect
of the melt pool aspect ratio and the initial (substrate) grain size
upon the selection of grain orientations and texture in the solidified melt pool.

\section{Methods}

While we explore a broader range of conditions, the models and
simulations used here are similar in nature to those already presented
and discussed in previous publications \cite{elahi2022multiscale,elahi2023grain}.
Therefore, here we only remind some of the main features of the modeling framework.
For more details, the reader is invited to refer to these prior works.

Essentially, we first calculate the alloy thermophysical data by
computational thermodynamics using the CALculation of PHAse
Diagram (CalPhaD) approach. Then, we perform the macro-scale
thermal analysis of SLM process to determine the temperature field
in and around the melt pool. Finally, we use the resulting thermal
profile and history to simulate the solidification microstructure
 with phase-field simulations at the micro-scale.
A one-way approach is used to link macro- and micro-scale
simulations.

\subsection{Computational alloy thermodynamics}

The temperature-dependent density, $\rho(T)$, enthalpy density,
$h(T)$, and effective heat capacity, $dh(T)/dT$ are computed by
the CalPhaD approach (software ThermoCalc, database TCNI8).
The phase diagram information is also computed by CalPhaD
using a dilute pseudo-binary approximation of Inconel 718 (IN718)
multi-component alloy. There are many different
ways to perform a pseudo-binary approximation of a
multi-component alloy \cite{raghavan2012construction}. Following
\cite{ghosh2017primary}, we consider Ni-5wt\%Nb as the
pseudo-binary surrogate for IN718. To make this approximation
consistent with the transformation temperatures of the full alloy, first we compute the equilibrium liquidus
temperature of IN718, $T_L$. Then, the partition coefficient of Nb, $k= c_s/c_l$,
in IN718 is calculated at $T_L$, with $c_s$ and $c_l$ the Nb concentration
in the solid and liquid phases, respectively. Finally, we calculate the slope of the
liquidus line of the pseudo-binary phase diagram, $m$, at $T_L$ as
$m = -dT_L/dc$ where $c$ denotes the concentration of
Nb in IN718. The rest of the phase diagram data is readily obtained
using the linear phase diagram assumption \cite{elahi2022multiscale}.

\subsection{Macroscopic thermal simulation}%
\label{sec:macro}%

Ignoring the effect of melt convection, the temperature history in
the melt pool is computed by the numerical solution of the
following heat transfer equation \cite{hong2019computer} in the
macro-scale spatial domain, $\Omega_M$,
\begin{equation}%
\label{eq:heat}%
\rho(T) \frac{dh(T)}{dT} \frac{\partial T}{\partial t} = \nabla
\cdot \big( K(T) \nabla T\big)
\end{equation}%
The temperature-dependent thermal conductivity, $K(T)$, is tabulated
from \cite{mills2002recommended}. The spatial domain, $\Omega_M$,
is a parallelepiped of dimensions $5\times2\times2$~mm$^3$ along x, y, and z
directions, respectively. 
Prior numerical convergence and size analysis \cite{elahi2022multiscale} 
revealed that the resulting thermal field for a single track in its quasi-steady state 
was essentially independent of the domain size beyond these dimensions.
The following boundary condition is
applied on the top surface of $\Omega_M$,
\begin{equation}%
\label{eq:heat:bc}%
- K(T) \nabla T\cdot \bm{n} = h_a (T-T_a) + \epsilon_R \sigma_R
(T^4 - T_a^4) + \frac{2\eta P}{\pi r_b^2}
\exp\bigg(\frac{2d(x,z,t)^2}{r_b^2}\bigg)
\end{equation}%
where $\bm{n}$ denotes the unit outward vector on boundaries of
$\Omega_M$, $h_a$ the convective heat transfer coefficient, $T_a$
the ambient temperature, $\epsilon_R$, the thermal radiation
coefficient, $\sigma_R$ the Stefan-Boltzmann constant, $P$ the
total laser power, $\eta$ the power absorption coefficient, $r_b$
the laser beam radius and $d$ denotes the distance to the beam
center. Assuming the initial location of the laser is denoted by
$(x_0, z_0)^T$ on the top surface, then for every point $(x, z)^T$
on this surface, $d$ is computed by $d(x,z,t) = \sqrt{(x-x_0-Vt)^2
+ (z-z_0)^2}$, where $V$ denotes the scanning velocity.
Importantly, neglecting fluid flow 
limits the applicability of our macroscopic analysis to
low laser input energy density, i.e. conduction mode, and it may 
not be as adequate for keyhole regime. Another effect of fluid flow
might be local temperature heterogeneities in 
the mushy zone, which are not represented with the elliptical temperature 
field approximation described further below.

Along the bottom surface, the powder bed is sitting on a substrate
with thermal properties of stainless steel
\cite{elahi2022multiscale} and its inferior surface at $T=T_a$.
The thermal properties of the powder bed are extrapolated from
that of the dense material, considering average particle diameter,
powder-bed compactness, inter-particle view factor, and properties
of the surrounding gas (Argon) \cite{elahi2022multiscale}. The
temperature-dependent properties of the material switch from
powder-bed to dense (fluid or solid) state when the local
temperature first exceeds the alloy liquidus temperature. A zero
heat flux condition is applied on the other boundaries. We solve
Eq.~(\ref{eq:heat}) by the finite element method with a
first-order Euler implicit scheme in time and trilinear hexahedral
elements in space, using a uniform spatial mesh of element size
$20~$\textmu m. Starting from the initial temperature $T_a$, the
simulation runs until the temperature field reaches to the quasi
steady-state conditions. Following
Ref.~\cite{elahi2022multiscale}, for our reference case, we used
$P=100~$W and $V=0.1~$m/s, with $r_b=35~$\textmu m,
$h_a=15~$W/m$^2$/K, $T_a=273~$K, $\epsilon_R=0.3$, and
$\eta=0.55$. %

\begin{figure}[!h]%
\centering{\includegraphics[width=\textwidth]{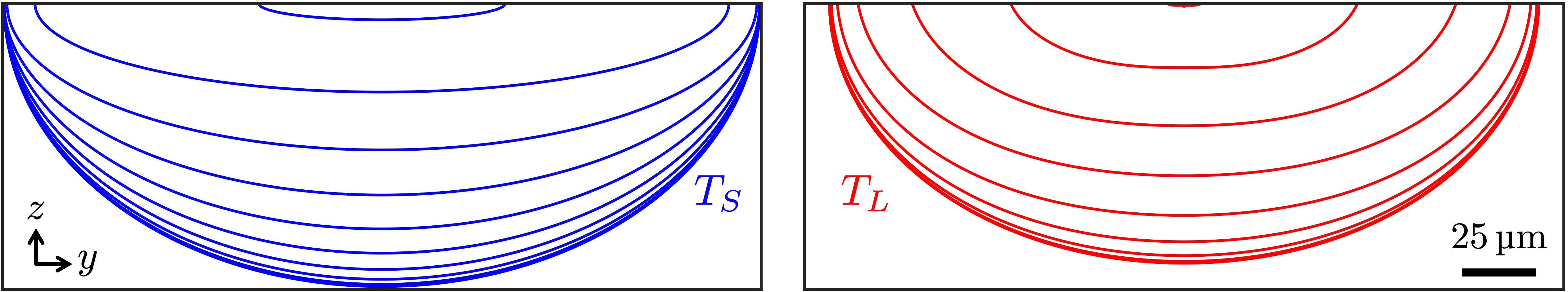}}%
\caption{
Time evolution of solidus (left) and liquidus (right)
isotherms in the $(y,z)$-plane showing the shrinking of the melt pool.
(Time increment between consecutive isotherms: 250~\textmu s.)
}%
\label{fig:temp}%
\end{figure}

Considering the location of the solidus isotherm, we denote $x_c$ and $x_e$
the $x$-coordinate of the deepest point and of the the tail of melt pool
with respect to the frame moving with the scanning velocity.
We then perform the two-dimensional phase-field simulation for $(y,z)$ cross-sections of
melt pool in the time interval $(0, t_f]$ where $t_f =
|x_c-x_e|/V$. For the sake of computational performance, we fit an analytical function to the
temperature field in the $(y,z)$-plane. For this purpose, the temperature
is interpolated radially between the solidus temperature and the temperature at
the center of melt pool $T(x_c,y_0,z_0)$. The solidus isotherm is approximated as
a half-ellipsis, and the central temperature is approximated using
analytical functions --- initially extrapolating radially to match reasonably the location of the
liquidus isotherm, then transitioning toward a constant cooling rate as the melt pool
center cools down below the liquidus temperature \cite{elahi2023grain}.
For the reference case, directly fitted to the finite element results,
the initial (maximal) half-depth and half-width of the solidus isotherm
are $d_S=96~$\textmu m and $w_S=128~$\textmu m, respectively.
Figure~\ref{fig:temp} shows the considered evolution of the elliptically approximated
solidus and liquidus isotherms in the $(y,z)$ cross-section as the melt
pool shrinks, considering these reference melt pool dimensions.
In order to study the effect of melt pool geometry on the
columnar grain growth competition, we simply multiply $(w_S,d_S)$
by factors $(0.5, 2.0)$, $(1.0, 1.0)$, and $(2.0, 0.5)$ for the deep,
reference, and wide melt pool scenarios, respectively.

\subsection{Microscopic grain growth simulation}%
\label{sec:micro}%

\subsubsection{Phase-field model:}%
\label{sec:micro:pf}%

We use a classical quantitative phase-field model for dilute binary alloys
solidification introduced in Ref.~\cite{echebarria2004quantitative},
with a non-linear preconditioning of the phase-field \cite{glasner2001nonlinear}
to reduce numerical mesh dependency.
The resulting form of the equations can be written
\begin{align}
\nonumber
\bigg(1- \frac{T-T_0}{mc_l^0}\bigg)\, %
a_s(\theta)^2\, \frac{\partial \psi}{\partial t} & = %
\nabla \big[ a_s(\theta)^2 \big]\cdot \nabla \psi + %
a_s(\theta)^2\, \bigg[ \nabla^2 \psi\, - \, \phi\, \sqrt{2}\,
|\nabla \psi|^2 \bigg] - %
\\ \nonumber & -
\frac{\partial}{\partial z}\, \bigg[ a_s(\theta)\,
a_s^\prime(\theta)\, \frac{\partial \psi}{\partial y } \bigg] + %
\frac{\partial}{\partial y}\, \bigg[ a_s(\theta)\,
a_s^\prime(\theta)\, \frac{\partial \psi}{\partial z } \bigg] %
\\ & +
\sqrt{2}\, \phi - \sqrt{2}\, \lambda (1-\phi^2)\, \bigg(U +
\frac{T(z,y,t)-T_0}{ mc_l^0(1-k) } \bigg)
\label{eq:P}%
\\\nonumber
\bigg(\frac{1+k}{2} - \frac{1-k}{2} \phi\bigg)\, %
\frac{\partial U}{\partial t} &= %
\nabla \cdot \bigg(\tilde{D}\, \frac{1-\phi}{2}\, \nabla U +
[1+(1-k)\, U]\, \frac{(1-\phi^2)}{4}\, \frac{\partial
\psi}{\partial
t}\, \frac{\nabla \psi}{|\nabla \psi|} \bigg) %
\\  &+
[1+(1-k)\, U]\, \frac{(1-\phi^2)}{2\sqrt{2}}\, \frac{\partial
\psi}{\partial t}%
\label{eq:U}%
\end{align}
where $T(z,y,t)$ denotes the temperature field that comes from the
macro-scale simulation results, $\phi$ the phase-field variable,
$\psi$ the preconditioned phase-field variable, $\phi(z,y,t) =
\tanh\big(\psi(z,y,t)/\sqrt{2}\big)$, $\theta =
\arctan{\big(\partial_y \psi/\partial_z \psi\big)}$ the
angle between the interface normal and the
horizontal axis, $U$ the dimensionless supersaturation, $U =
\frac{1}{1-k}\, \big(\frac{2\, c/c_l^0}{1\,-\,\phi\, +\,
k\,(1\,+\, \phi\, )}\, -\, 1\big)$, and $c$ the solute concentration
field, $c^0_l = c_\infty/k$ the solute concentration of a flat
interface at the reference (solidus) temperature $T_0$ for an
alloy of nominal solute concentration $c_\infty$.
In Eqs~\eqref{eq:P}-\eqref{eq:U}, space is scaled in units of
the diffuse interface width, $W$, and time is
scaled in units of the relaxation time, $\tau_0$, at
$T_0$ \cite{echebarria2004quantitative}. The capillarity length, $d_0$,
is computed at $T_0$
\cite{echebarria2004quantitative}, $d_0 = \Gamma/\big(|m|c_\infty
(1/k - 1)\big)$, where $\Gamma$ denotes the Gibbs-Thomson
coefficient of the solid-liquid interface. The non-dimensional
value for the liquid diffusion coefficient, $\tilde{D}$, and the
coupling factor, $\lambda$, are computed according to the
following identities: $\tilde{D} = D\tau_0/W^2 = a_1 \,
a_2 \, W/d_0$, $\lambda = a_1\, W/d_0$, where $D$
denotes the liquid diffusion coefficient, $a_1 = 5\sqrt{2}/8$ and
$a_2 = 47/75$.
We use a standard form of the fourfold anisotropy of the surface
tension $\gamma(\bar{\theta}) = \bar{\gamma} a_s(\bar{\theta})$, with $a_s(\bar{\theta}) = 1+ \epsilon_4\,
\cos\,(4\bar{\theta})$, where $\bar{\gamma}$ is the average
surface tension in a $\langle 1 0 \rangle$ plane, $\epsilon_4$ is
the strength of the surface tension anisotropy, and $\bar{\theta}$
is the angle between the normal to the interface and a fixed
crystalline axis. For a crystal misorientation $\alpha_0$ with
respect to the coordinate axes, the anisotropy as function of
$\theta$ between the interface normal and the $x$-axis follows
$a_s(\theta) = 1 + \epsilon_4\, \cos\,\big(4\, (\theta - \alpha_0)\big)$.
As the considered laser speed is relatively low ($V=0.1~$m/s), the solid-liquid
interface is assumed to remain under local equilibrium conditions.
Therefore, the model does not consider kinetic undercooling or solute
trapping, such that $\tau_0$ is computed as $\tau_0 =
a_2\, \lambda\, W^2/D$, and $W$ is the only model
parameter that should be appropriately chosen for the purpose of
quantitative prediction.
As in Refs~\cite{elahi2022multiscale,elahi2023grain}, the material properties for the Ni-Nb (IN718 surrogate) alloy used in PF simulations are
$c_\infty=5.0\,$wt\%Nb,
$k=0.37$,
$m=9.0\,$K/wt\%Nb,
$\Gamma=2.49\times10^{-7}~$Km,
$D=2.44\times10^{-9}~$m$^2$/s,
$\epsilon_4=0.02$,
with
$T_M=1670.43~$K and
$T_0=1548.81~$K.

We solve Eqs.~\eqref{eq:P}-\eqref{eq:U} under homogeneous Neumann (no-flux)
boundary conditions along all directions. Moreover, the
solid-liquid interface is initialized along the liquidus isotherm, with
$\psi$ initialized as the signed distance function to the
liquidus isotherm (negative in the liquid region). The
dimensionless supersaturation field is initialized based on the
equilibrium concentration, i.e. $U(z,y,0) = -1.0$. The
computational domain $\Omega_m$ is a rectangle of dimensions $2.01
w_s \times 1.01 d_s$, e.g. $\Omega_m = 257~$\textmu m$\,\times\,
97~$\textmu m in the case of the reference melt pool configuration, with the
center of the melt pool at the center of the top boundary (as represented in Fig.~\ref{fig:temp}). The
total simulated time is equal to 2.5~milliseconds in all cases.

\subsubsection{Polycrystalline grain growth:}%
\label{sec:micro:poly}%

To consider the columnar grain growth competition in the melt
pool, the two-dimensional (2D) orientation of grains is stored in an
auxiliary integer field. It assumes a value of $0$ in the
liquid phase and an integer in the range of $[1, 90]$ in the solid
phase, which is also used as a discrete set of misorientation angles of the grains,
in degrees.
When $(1-\phi^2)$ exceeds a threshold,
here $0.01$, the grain index is updated according to the
local neighborhood. This method creates a halo
of orientation field in the liquid around a grain. When a grid point is
allocated a solid grain index value higher than $0$, the
index field no longer evolves, and neither does the solid-solid grain boundary (GB).
While it does not account for solid-state microstructure evolution, this method is a computationally efficient
alternative to multi phase-field models \cite{steinbach1996phase},
since it relies on a single phase field.
Its use is appropriate in the presence of well-developed dendritic structures,
with GBs forming deep within the mushy zone, i.e. when the macroscopic orientation
of resulting GBs depends primarily on the growth competition of primary and secondary dendrite tips
in the the vicinity of the solidification front.

To initialize the grain index field, N$_g$ number of points
(Voronoi cell centers) are randomly distributed in $\Omega_m$. A
random integer value in the range of $[1, 90]$ is attributed to
each point as the corresponding grain orientation. After the spatial
discretization of $\Omega_m$, the initial grain distribution is generated by a
classic Voronoi tessellation algorithm. Finally, the grain orientation of
computational cells located within the liquid region ($\psi <
0$) are reset to $0$.
The solution of Eqs~\eqref{eq:P}-\eqref{eq:U} hence results in the
epitaxial growth of columnar grains from the melt pool fusion line.
Since we aim to assess the effect of initial grain size on the solidification microstructure, we
considered different grain densities with N$_g$ = 750, 1500, 3000 and 6000, corresponding
to equivalent grain radii between 1.15 and 3.25\,\textmu m.
Moreover, in order to gain a better statistical picture of final grain
structure, each simulation is repeated five times with different
initial grain distributions, hence resulting in 60 PF simulations in total.

\subsubsection{Implementation:}%
\label{sec:micro:imp}%

Phase-field Eqs~\eqref{eq:P}-\eqref{eq:U} are solved in
two dimensions by the finite difference method on a uniform
spatiotemporal grid using the Euler explicit time integration scheme.
The time step size, $\Delta t$,  is considered as $0.3$ of the
maximum time step size based on the stability of Laplacian
operators. A standard second-order five-point stencil is used to
discretize Laplacian operators. The rest of terms in Eqs~\eqref{eq:P}
and \eqref{eq:U} are discretized by central difference schemes
(see appendices of Ref.~\cite{tourret2015growth} for details).

The diffuse interface width $W = 0.8\, \Delta x$ and grid spacing
$\Delta x$ were determined based on a convergence study of the
steady-state tip undercooling as a function of grid size in a
unidirectional solidification with temperature gradient equal to
its average in the mushy region and a pulling velocity equal to
the scanning speed. Under conditions relevant to additive
manufacturing, such a convergence study is quite limiting, since
both the dendrite tip radius and the diffusion length are small.
However, it is essential for the purpose of quantitative
prediction of dendrite/cells growth kinetics and resulting grain
structures. While the convergence study pointed at $\Delta
x\approx 5~$nm for well-converged simulation in longitudinal
simulations \cite{elahi2022multiscale}, here we can afford to use
$\Delta x=10~$nm, since isotherms and interface velocities are
lower along the cross-section than they are along the
longitudinal direction. Hence, the total number of spatial
computational cells is equal to $25700 \times 9700 \approx 2.49
\times 10^8$ grid points in the case of the reference melt pool
dimensions. While $(y,z)$ dimensions are different, the total
number of grid points is the same for the deep and wide
configurations. The resulting time step is equal to $3\times
10^{-9}$~seconds, such that O($10^{5}$) time steps are required to
complete the simulation.

Because of this grid size limitation for quantitative
predictions, PF simulations at the scale of a full melt pool,
even in 2D, are extremely computationally demanding.
Therefore, advanced acceleration schemes are required.
We implemented the model for massively parallel computing on
multi-graphic processing units (Multi-GPU) with the computer
unified device architecture (CUDA) programming language.
Each simulation is performed on one computing node equipped with four Nvidia RTX-3090 GPUs.
We use a layer-wise domain decomposition to distribute computation
load among GPUs, with the computational domain divided
into 4 almost equal layers along the $y$-direction (and the same
number of grid points along $x$). An extra halo grid layer
is added to the top and bottom rows of each domain to simplify the
imposition of boundary conditions and inter-GPU data
communication. Halo layer data is updated using direct
GPU-GPU communication, so as to avoid expensive
GPU-to-CPU and CPU-to-GPU data transfers.

\section{Results and Discussion}

\begin{figure}[!t]%
\centering
\includegraphics[width=.8\textwidth]{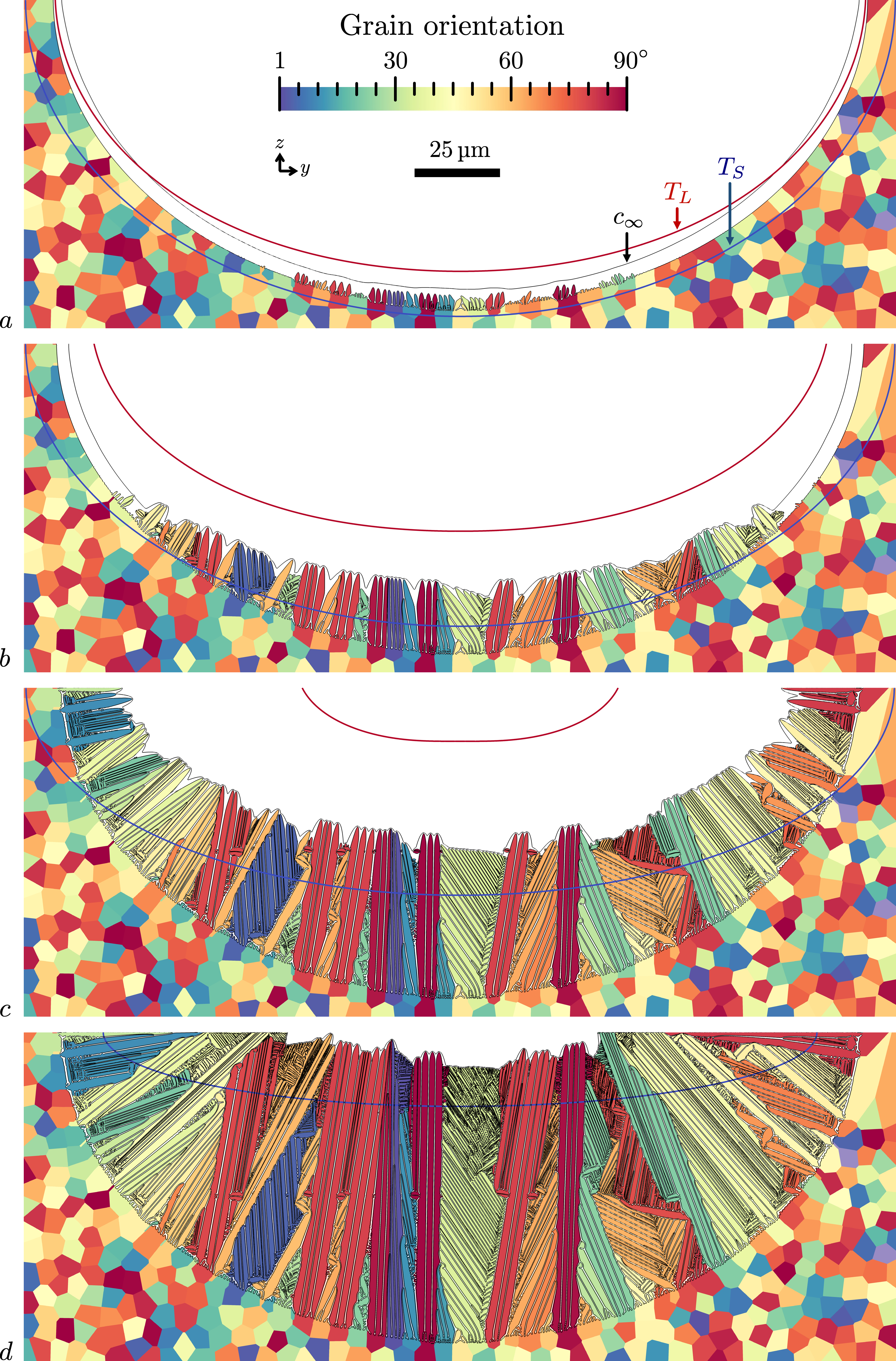}
\caption{
Evolution of grain structure with time for the reference melt pool size
and N$_g$ = 750, at $t = $(a) 0.75, (b) 1.25, (c) 1.75, and (d) 2.25~ms.
Solidus and liquidus isotherms are marked in blue and red line, respectively, and the iso-concentration $c=c_\infty$ appears as black line.
\label{fig:grain_evol}
}%
\end{figure}

\begin{landscape}

\begin{figure}[!t]%
\includegraphics[width=\columnwidth]{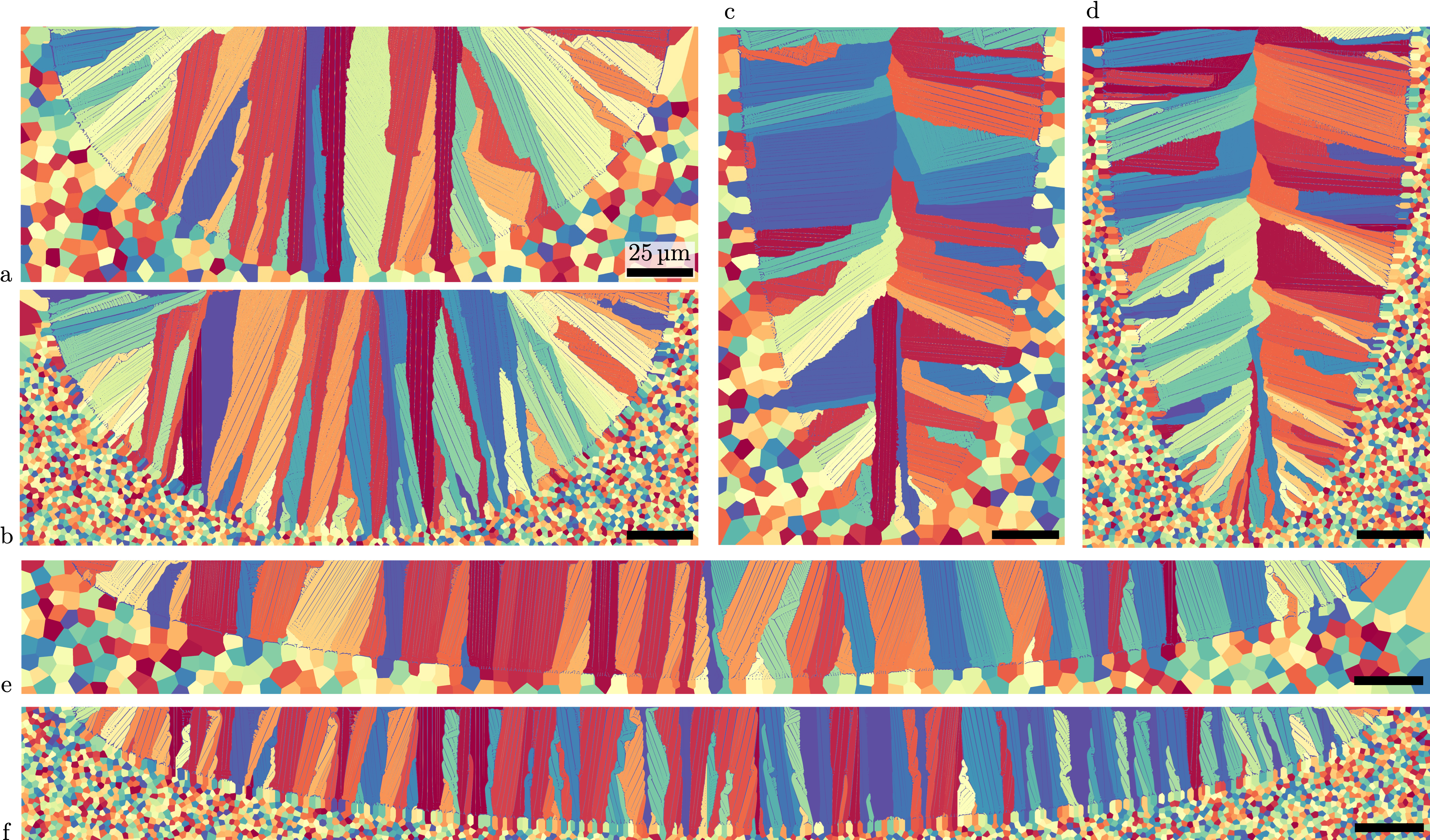}%
\caption{Representative final grain structures corresponding to reference (a,b),
deep (c,d) and wide (e,f) melt pools with N$_g$ = 750 (a, c, e) and N$_g$ = 6000 (b, d, f).
The grain orientation color map is similar to Fig.~\ref{fig:grain_evol}.
}%
\label{fig:grain_final}%
\end{figure}%

\end{landscape}

Figure~\ref{fig:grain_evol} illustrates the evolution of the grain
structure during competitive growth in a simulation
for the reference melt pool size and a grain number N$_g$ = 750.
Figure~\ref{fig:grain_final} shows the final grain structures for
one of the five simulation for each of
the three considered melt pool sizes (reference, deep, and wide)
and two considered initial grain numbers (N$_g$ = 750 or
6000).

In all cases, the solid-liquid interface initially advances with a locally
nearly-planar morphology at early stage. After a short
time period, required for the development of a concentration boundary layer
ahead of the interface, it destabilizes into cellular patterns.
Several cells are eliminated early, leading to a rapid increase of the local
primary spacing, while dendritic sidebranches appear after a short time.
The average primary dendritic arm spacing (PDAS) was not found to markedly
depend upon the initial grain density.
Moreover, the iso-concentration line at $c=c_\infty$ in Fig.~\ref{fig:grain_evol} (black line), which closely envelops the solidification front,
shows that the diffusion length is of the same order or even smaller than the selected local primary spacings.
While the diffusive conditions are an oversimplification of the transport regime in the melt pool
--- strong Marangoni convection is usually expected --- this may provide an explanation
for the typical lack of secondary sidebranches in additively manufactured metallic alloys.

A strong growth competition occurs between columnar grain in the
melt pool. In qualitative agreement with the theory by Walton and
Chalmers \cite{walton1959origin}, grains with a $\langle10\rangle$
crystalline axis well aligned with the main heat transfer
directions (i.e. perpendicular to the fusion line and isotherms)
tend to prevail in the growth competition by eliminating the less
favored grains with higher misorientation with the temperature
gradient. In the case of the reference melt pool, the melt
pool aspect ratio is close to unity (i.e. the fusion line is near
circular). The resulting growth competition does not lead to a
strong texture, as seen by the broad distribution of orientations
present in the solidified melt pool
(Fig.~\ref{fig:grain_final}\,a,b). In contrast, in the case of deep
(Fig.~\ref{fig:grain_final}\,c,d) and wide
(Fig.~\ref{fig:grain_final}\,e,f) melt pools, the growth
competition results in a more noticeable $\langle10\rangle$ texture
(apparent from the predominance of darker shades of red and blue).
This $\langle10\rangle$ texture is primarily attributed to the
large fraction of the melt pool solidifying  under a mostly
horizontal (Fig.~\ref{fig:grain_final}\,c,d) or vertical
(Fig.~\ref{fig:grain_final}\,e,f) temperature gradient. The grain
elimination mechanism proceeds via either dendrite impingement in the
case of converging dendrites (grains) or side-branching in the
case of diverging dendrites \cite{tourret2015growth}.

\begin{figure}[!ht]%
\includegraphics[width=\textwidth]{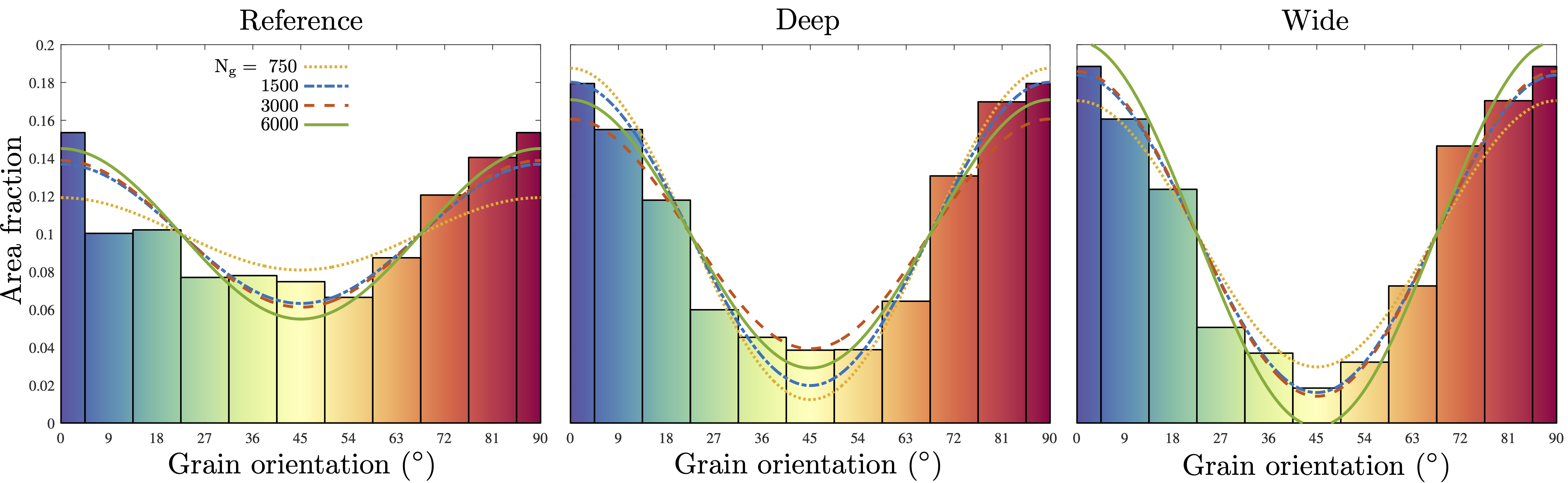}%
\caption{Histograms of grain orientations for reference (left),
deep (center), and wide (right) melt pools. Each histogram shows the
orientation distribution combining all four initial grain densities (20 simulations per histogram).
Overlapped curves are the fitted cosine functions for the 4 different initial grain densities (5 simulations each).
The grain orientation distributions are counted over each solid pixel and
scaled by the total number of counted grid points, such that the ordinate
axis corresponds to a fraction of the the melt pool area.
Considering the fourfold symmetry of the crystal structure, histograms are periodic with respect to the orientation angle, such that the leftmost and rightmost bins are similar and include angles from $-4^\circ$ (i.e. $86^\circ$) to $4^\circ$ (i.e. $94^\circ$).
}%
\label{fig:hist}%
\end{figure}

To further analyze texture selection, we quantified the
orientation distributions in the different simulations.
In Figure~\ref{fig:hist}, histograms show normalized
grain orientation (i.e. fraction of the melted area) with $N_{\rm bin}=10$ bins of width $9^\circ$,
combining all grain densities (20 simulations per histogram).
To illustrate the dependence upon initial grain densities,
additional curves show idealized functions $H(\theta)=A\cos(4\theta)+1/N_{\rm bin}$,
fitted to similar orientation distributions histograms within the five runs for each
different initial grain density. The sole fitting parameter $A$ may be
interpreted as a measure of the strength of the $\langle10\rangle$ texture,
with $A=0$ corresponding to a flat distribution (no texture).
This analysis confirm that deep ($A\approx0.0749$) and wide ($A\approx0.0855$) melt pools have a strong
$\langle10\rangle$ texture (high population of grains with
near 0 and 90 degrees orientations) in comparison to the
reference melt pool geometry ($A\approx0.0349$).
Fitted curves for different N$_g$ show that the effect of initial grain size on the final texture is marginal, with most values of $A$ (all values for deep and wide cases) deviating by 20\% or less from those obtained for the distributions combining all values of N$_g$ (listed in parenthesis earlier).

In the case of a deep melt pool, we can essentially divide the melt pool
into two regions: a relatively texture-less bottom part with a circular/elliptical fusion line, and a textured upper part with a near
vertical fusion line. As a result, the final $\langle10\rangle$ texture in the
deep case is less strong compared to the wide melt pool,
in which the main thermal gradient direction is essentially vertical.
Interestingly, while these results show a clear texture transition
from weak to strong $\langle10\rangle$ texture when the melt pool aspect
ratio deviates from unity, they do not exhibit any indication of a transition
toward other textures --- such as, for instance, a possible $\langle110\rangle$ texture
suggested from geometrical arguments and experimental hints \cite{jadhav2019influence,higashi2020selective}.

Finally, on the computational side, a preliminary scaling analysis shows a near linear scaling of the algorithm
speedup with the number of GPUs. For instance, for the reference melt pool and $N_g = 6000$, the
computational time is about $82$, $43$ and $22$ hours using 1, 2 and 4 GPUs, respectively.
Still, while the multi-GPU parallelization goes a long way in enabling full melt pool scale simulation in 2D, a similar 3D investigation remains unreasonable.
Indeed, a single equivalent three-dimensional quantitative PF simulation at full melt pool scale would require a
number of spatial grid points of O($10^{13}$) and a similar number of time steps.
Such a large-scale simulation, even using state-of-the-art petascale computing resources,
would require O(day) to perform.
It would therefore result in an enormous investment in time and resources for a relatively minor return in terms of physical insight into grain texture selection.
For 3D simulations at this scale, coarse-grained approaches, e.g. cellular automaton, remain most convenient, while statistically capturing grain texture selection with reasonable accuracy \cite{elahi2023grain}.

\section{Summary and Perspectives}

We performed quantitative phase-field simulation of solidification
in thermal conditions relevant to SLM processing of a Nickel alloy.
We limited our micro-scale simulations to two-dimensional
cross-sections of the melt pool (perpendicular to the
scanning direction), using a thermal history at quasi
steady-state computed by 3D finite element analysis, using
CalPhaD-computed alloy parameters \cite{elahi2022multiscale,elahi2023grain}.
We studied the effect of the melt pool geometry (i.e. its aspect ratio)
and of the initial grain density along the fusion line on the
final microstructure. According to our results, a near-circular fusion line
(i.e. an aspect ratio close to unity) does not lead to any notable texture,
while relatively wider and/or deeper melt pools, which have more
horizontally or vertically oriented fusion lines, promote the formation of a clear
$\langle10\rangle$ texture.

Here, we only considered the single-track melting of a random Voronoi 
grain distribution. Therefore, the resulting grain texture is not completely
representative of a realistic AM microstructure. Indeed, we decided 
to put the focus on the fundamental mechanisms of grain growth 
competition within a non-one-dimensional temperature field, which  
have received little attention so far.
In order to produce a more realistic simulation of grain texture 
emergence from AM, one should account for multiple layers, 
multiple tracks, and nucleation events, 
all of which would be relatively straightforward to implement, as well as 
three-dimensional simulations, which, on the other hand, could 
be much more challenging computationally.

Perspectives following from this work are multiple. A
three-dimensional quantitative phase-field study with the same
level of accuracy (down to the level of individual dendrite)
remains computationally prohibitive, and is thus unlikely to
provide statistically-meaningful insight. However, even in two
dimensions, the inclusion of additional physics into the
simulations could certainly bring a fresh look onto grain growth
competition within multidimensional temperature fields. For
instance, including crystal nucleation in the melt pool would make
it possible to study potential columnar-to-equiaxed transitions.
The present study was also limited to the lower end of the
velocity range relevant to SLM processing. In this low-velocity
regime, the solid-liquid interface remains close to equilibrium,
which is not always the case for typical SLM conditions. Ongoing
work specifically focus on extending this kind of study to include
kinetic undercooling and solute trapping using recent PF
formulations \cite{pinomaa2019quantitative,ji2023micro}.
%

\ack
This work was supported by the Spanish Ministry of Science and the European Union NextGenerationEU (PRTR) through the MiMMoSA project (PCI2021-122023-2B) and a Ram\'on y Cajal fellowship (RYC2019-028233-I).

\section*{References}

\bibliographystyle{iopart-num} 
\bibliography{biblio}%

\end{document}